\documentclass{acm_proc_article-sp}
\usepackage{amsmath}
\usepackage{dsfont}
\usepackage{times}
\usepackage{helvet}
\usepackage{courier}
\usepackage{graphicx}
\newcommand{\remove}[1]{}

\newcommand{\figref}[1]{Figure~\ref{#1}}
\begin{document}

\title{Structural and Cognitive Bottlenecks to Information Access in Social Networks}
\numberofauthors{2}
\author{
\alignauthor
Jeon-Hyung Kang \\
       \affaddr{USC Information Sciences Institute}\\
       \affaddr{4676 Admiralty Way}\\
       \affaddr{Marina del Rey, CA 90292}\\
       \email{jeonhyuk@isi.edu}
\alignauthor
Kristina Lerman\\
       \affaddr{USC Information Sciences Institute}\\
       \affaddr{4676 Admiralty Way}\\
       \affaddr{Marina del Rey, CA 90292}\\
       \email{lerman@isi.edu}
}

\toappear{Permission to make digital or hard copies of all or part of this work for personal or classroom use is granted without fee provided
that copies are not made or distributed for profit or commercial advantage and that copies bear this notice and the full citation on the
first page. To copy otherwise, or republish, to post on servers or to redistribute to lists, requires prior specific permission and/or a
fee.\\ \emph{\small{24th ACM Conference on Hypertext and Social Media\\ 1--3 May 2013, Paris, France}}\\ Copyright 2013 ACM }
\maketitle
\begin{abstract}
Information in networks is non-uniformly distributed, enabling individuals in certain network positions to get preferential access to information. Social scientists have developed influential theories about the role of network structure in information access. These theories were validated through numerous studies, which examined how individuals leverage their social networks for competitive advantage, such as a new job or higher compensation.
It is not clear how these theories generalize to online networks, which differ from real-world social networks in important respects, including asymmetry of social links. We address this problem by analyzing how users of the social news aggregator Digg adopt stories recommended by friends, i.e., users they follow. We measure the impact different factors, such as network position and activity rate; have on access to novel information, which in Digg's case means set of distinct news stories. We show that a user can improve his information access by linking to active users, though this becomes less effective as the number of friends, or their activity, grows due to structural network constraints. These constraints arise because users in structurally diverse position within the follower graph have topically diverse interests from their friends. Moreover, though in most cases user's friends are exposed to almost all the information available in the network, after they make their recommendations, the user sees only a small fraction of the available information.
Our study suggests that cognitive and structural bottlenecks limit access to novel information in online social networks.

\end{abstract}


\section{Introduction}
Sociability confers survival advantage~\cite{Dunbar03,Silk07}. As a result, humans have evolved a set of specialized skills for maintaining a large number of  complex social connections~\cite{Dunbar,Herrmann07}. They use these skills daily to link to other individuals and then exploit the resulting social network for personal advantage.
Social scientists have long been interested in the role of network structure in the performance of individuals and organizations. A pair of classic theories has linked position within a network to successful outcomes for individuals. Burt~\cite{Burt95,Burt04} attributed individual success (better job outcomes, higher compensation) to their brokerage positions within the network. These positions link distinct communities, thereby exposing brokers to diverse and novel information that others within their community may not have. Granovetter~\cite{granovetter1973} similarly argued that novel information tends to flow to us via weak ties, i.e., acquaintances from other communities with whom we interact infrequently. Our strong ties, on the other hand, are close friends who belong to the same community, and therefore have the same information we do. Aral \& Van Alstyne~\cite{Aral11} linked the two theories, arguing that positions of greater network diversity, i.e., linking to others who are not otherwise connected, are associated with lower rate of interactions (low channel bandwidth).
They further demonstrated, by analyzing a corpus of email communications, that this trade-off limits the amount of diverse and novel information that individuals in brokerage positions receive, though they did not speculate as to the origin of this trade-off.

Recently, social media has emerged as an important platform for social interaction and information exchange. On social media sites, such as Twitter and Digg, users create social networks by subscribing to, or following, other users. When a user posts a message, or a hyperlink to content they found online, this message is broadcast to all followers, who may themselves choose to share it with their own followers, and so on, enabling the message to spread over the social network.

Social media provides us with new data for testing and generalizing information brokerage theories. In this paper we study the interplay between network structure, user activity and information content. Despite significant differences between online social networks and the email networks studied by Aral \& Van Alstyne, we validate the main conclusion of their study, namely the trade-off between network structure and channel bandwidth. We find that the diversity of information that social media users are exposed to via friends depends on their position in the network. Users embedded within a community of strongly tied individuals are likely to share information on topics that other community members are interested in, while users in brokerage  positions that bridge different communities receive information on more diverse topics. Users can increase their access to novel information by adding more friends. However, by adding friends, they also increase the quantity of information they are exposed, often beyond their capacity to process it~\cite{Hodas12socialcom}. Furthermore, increasing the quantity of information does not necessarily increase its novelty or diversity, since network structure has important implications for information diversity.

The paper makes the following contributions. In Section~\ref{sec:data}, we describe the data we use from the social news aggregator Digg, and define a set of network and information variables we use to characterize access to information in this network. In Section~\ref{sec:networkstructure}, we investigate the relation between the structure of the social network, the content of information, and the activity of users and their friends.
We study how the amount of novel information available to the user, specifically, the number of distinct news stories, depends on network structure and user activity.
We test the existence of trade-off between network diversity and friends' activities in online social networks.
Our study suggests that cognitive and structural bottlenecks limit access to novel information in online social networks.

\section{Data and Methods}
\label{sec:data}

Social news aggregator Digg allows registered users to submit links to news stories and other users to vote for the stories they find interesting. Digg also allows users to follow the activity of other users. The follow links are not necessarily reciprocated: a user $b$ who follows user $a$, can see the messages $a$ posts, but not vice versa. We refer to $a$ as the \emph{friend} of $b$, and $b$ as the \emph{follower} of $a$. A user's social stream shows the stories his friends submitted or voted for. When the user votes for any story, this vote is broadcast to all of his followers who can themselves see it in their social streams.

At the time datasets were collected, users were submitting tens of thousands of stories, from which Digg selected a handful (about 100 each day) to promote to its front page based on how popular the story was in the community. Before a story is promoted to the front page, it is visible on the upcoming stories queue and to submitter's followers via their social stream (friends' interface). With each new vote, the story becomes visible to the voter's followers.

\subsection{Data processing}
We analyzed two data sets collected in the past from Digg.
The 2009 data set~\cite{Lerman10icwsm} contains information about the voting histories of 3.5K stories promoted to Digg front page in June 2009,
and contains 2.1 million votes by 70K users. The follower graph of these voters contains 1.7 million social links. At the time this dataset was collected, Digg was assigning stories to one of 8 topics (Entertainment, Lifestyle, Science, Technology, World \& Business, Sports, Offbeat, and Gaming) and one of 50 subtopics (World News, Tech Industry News, General Sciences, Odd Stuff, Movies, Business, World News, Politics, etc.).
The 2010 data set~\cite{sharara:icwsm11}  contains information about voting histories of 11,942 users over a six months period (Jul - Dec 2010). It includes 48,554 stories with 1.9 million votes. The follower graph contains 1.3 million social links. At the time data was collected, Digg assigned stories to 10 topics (Entertainment, Lifestyle, Technology, World News, Offbeat, Business, Sports, Politics, Gaming and Science) replacing the 2009 ``World \& Business'' category with ``World News,'' ``Business,'' and ``Politics''.

We examine only the votes that the story accrues before promotion to the front page.  During that time, it propagates mainly via friends' recommendations. After promotion, users are likely to be exposed to the story through the front page, and vote for it independently of friends' recommendations. In the 2009 data set, 28K users voted for 3,553 stories and in the 2010 data set, 4K users voted for 36,883 stories before promotion. We focused the data further by selecting only those users who voted at least 10 times, resulting in 2,390 users (who voted for 3,553 stories) in the 2009 data set and 2,330 users (who voted on 22,483 stories) in the 2010 data set.

\subsection{Definition of Variables}
Following Aral \& Van Alstyne~\cite{Aral11,aral2012anatomy} we define a set of variables to characterize access to information in networks.
\remove{
\begin{table*}
\center
\begin{tabular}{|l|l|c|}
\hline
\textbf{Symbol} &\textbf{   Variable }  & \textbf{Definition} \\ \hline
$S_i$ & num. active friends & $S_i=\sum_{u_j \in N^{frd}_i}{\delta(u_j)}$\\
$ND_i$ & network diversity & $ND_i=1-\frac{ |\{ e_{jk}: u_j, u_k \in N_i, e_{jk} \in E \} |} { |N_i| (|N_i| -1)} $ \\
$O_i$ & volume of outgoing info. & num. initiations ($O_{i}^s$) + num. adoptions ($O_{i}^a$) made by $u_i$  \\
$I_i$ & volume of incoming info. &  $I_i=\sum_{k=1}^{N^{frd}_i}O_k$\\
$B_i$ & avg friend activity& $B_{i}= \frac{I_{i}}{S_{i}}$\\
$uB_i$ & num. adoptions & $uB_{i}= O_{i}^a$\\
$TD_i$ & friend topic diversity & $TD_{i}= \frac{ \sum_{j=1}^{N^{frd}_i} \sum_{k=1}^{N^{frd}_i} (1-Cos(\theta_{jt},\theta_{k}))}{S^{2}_{i}}
$\\
$NRI_i$ & novel information &  num. distinct stories received\\
$NRI^{frds}_i$ & novel info. potential & num. distinct stories received by friends of user $i$ \\
$R_i$ & novel info. rate & $R_i=NRI_i/I_i$ \\
$NAR_{i}$ & novel info. adoption rates & $NAR_{i} = O_{i}^a/NRI_{i}$ \\
$FNAR_{i} $ & friend novel info. adoption rates & $FNAR_{i} = NRI_{i}/NRI^{frds}_i$ \\
\hline
\end{tabular}
\center
\caption{Variables used in the study and their definitions.}
\label{tbl:definitions}
\end{table*}
}
\begin{table}
\center
\begin{tabular}{|l|l|}
\hline
\textbf{Variable} &\textbf{   Description }   \\ \hline
$S$ & number of active friends \\ 
$ND$ & network diversity \\ \hline 
$O$ & volume of outgoing info. (\# votes by user) \\
$I$ & volume incoming info. (friend recommendations)\\
$B$ & avg friend activity \\ 
$uB$ & user activity (\# adopted recommendations) \\ 
$TD$ & friend topic diversity \\ \hline 
$NRI$ & novel information \\
$NRI^{frds}$ & novel information friends are exposed to \\
$NAR$ & fraction of novel information adopted by user \\
$FNAR $ & fraction of novel information adopted by friends \\
\hline
\end{tabular}
\center
\caption{Variables used in the study.}
\label{tbl:definitions}
\end{table}

\subsubsection{Network Variables}
A social network can be represented as a graph $G = (U,E)$ consisting of a set of users $U$ and a set of edges $E$ between them. There exists edge $e_{ij} \in E$, if user $u_i$ follows user $u_j$. While in traditional social networks, friendship links as reciprocated, resulting in an undirected graph, online social networks (e.g., Twitter and Digg) form a directed graph. It allows users to follow people with certain interests without having a reciprocal relationship.
The neighborhood $N_i$ of user $u_i$ consists of both friends $N^{frd}_i$ and followers $N^{fol}_i$ of $u_i$.

\paragraph{Network Size}
Network size is an important variable which shows the breadth of contacts each user has.  We define the size of $u_i$'s network, $S_i$, as the number of friends from whom user $u_i$ received messages during certain time period $\Delta T$, which we take to be the time over which data was collected. Since not all friends were active during that period and thus had a chance to influence $u_i$ votes, we focused on active friends, i.e., friends who had recommended stories during $\Delta T$.  Therefore, network size is defined as
 \begin{equation}
\begin{aligned}
S_i=\sum_{u_j \in N^{frd}_i}{\delta(u_j)}
 \end{aligned}
\end{equation}
\noindent where $\delta(u_j)$ is one if and only if $u_j$ voted at least ten times during the time period $\Delta T$ and zero otherwise. Note that ten is the minimum number of messages to cover all topic categories in 2009 and 2010 data set.

\paragraph{Network Diversity}
Network diversity of user $u_i$ represents how many otherwise unconnected neighbors $u_i$ interacts with.
We measure network diversity using local clustering coefficient \cite{watts1998small}, $C_i$, which quantifies how often the neighbors of $u_i$ are linked together (regardless of the direction of the edge):
 \begin{equation}
\begin{aligned}
{C_i= \frac{ |\{ e_{jk}: u_j, u_k \in N_i, e_{jk} \in E \} |} { |N_i| (|N_i| -1)} }
 \end{aligned}
\end{equation}
\noindent where $N_i$ is the set of neighbors of user $u_i$ and $|N_i|$ is the number of neighbors. The total number of possible connections among neighbors is $|N_i| (|N_i|-1)$. High clustering coefficient implies low network diversity, and vice versa. Therefore, we define network diversity of user $u_i$ as $ND_i=1-C_i$.
Aral \& Van Alstyne~\cite{Aral11,aral2012anatomy} defined network diversity as the lack of structural holes using the first and second order dimensions of link redundancy. We prefer to follow the definition of Watts et al.~\cite{watts1998small}, since clustering coefficients are more evenly distributed over the range from 0 to 1.

\subsubsection{User Activity Variables}
Access to information in social network depends on the activity levels of users. In friendship networks, strength of a tie defines frequency and intensity of interaction of a pair of individuals~\cite{granovetter1973}. Close friends --- strong ties --- interact more frequently than acquaintances (weak ties). In the analysis of email communications, Aral \& Van Alstyne used the quantity \emph{channel bandwidth} to represent the strength of a tie. They defined bandwidth as the number of messages sent across the tie. One-to-many directed broadcasts of social media differ in nature from email communication. We find it useful to separate activities into incoming messages and outgoing messages.

\paragraph{Average Friend Activity} 
In social media, friends' activity determines the total volume of incoming information $I_{i}$  over a time period $\Delta T$. We measure $I_i$ as the total number of stories friends of user $u_i$ recommended, i.e., voted for, during the time period $\Delta T$. Hence, we define the average (per link) volume of incoming information during $\Delta T$ as:
 \begin{equation}
\begin{aligned}
{B_{i}= \frac{I_{i}}{S_{i}} }
 \end{aligned}
\end{equation}

\paragraph{User Activity}
Most social media sites, including Digg and Twitter, display items from friends as a chronologically sorted list, with the newest items at the top of the list. A user scans the list and if he finds an item interesting, he may share it with his followers, e.g., by voting it. He will continue scanning the list until he loses interest, gets bored or distracted~\cite{Hodas12socialcom}. When user gets bored, he could start to inspect new  information from outside of his social network and recommend new information to all his followers. User $u_i$'s activity is the sum of the number of new stories  $O_{i}^s$ (seeded messages) user discovered from outside of his network by browsing the Web and other sections of Digg, and the number of stories $O_{i}^a$ he adopted from friends' recommendations.
In the analysis presented in this paper we focus on the component of user activity that corresponds to adoption events, i.e., cases where user votes for a story after a friend had recommended it. Therefore, we measure the activity of the user $u_i$ as the number of adoptions the user made during the time period $\Delta T$:
 \begin{equation}
\begin{aligned}
{uB_{i}= O_{i}^a} 
 \end{aligned}
\end{equation}

\subsubsection{Information Variables}

We model information content in a user's network using topic diversity of information and the total volume of novel information.

\begin{figure}[]
\begin{center}
\begin{tabular}{c}
\includegraphics[width=0.95\linewidth]{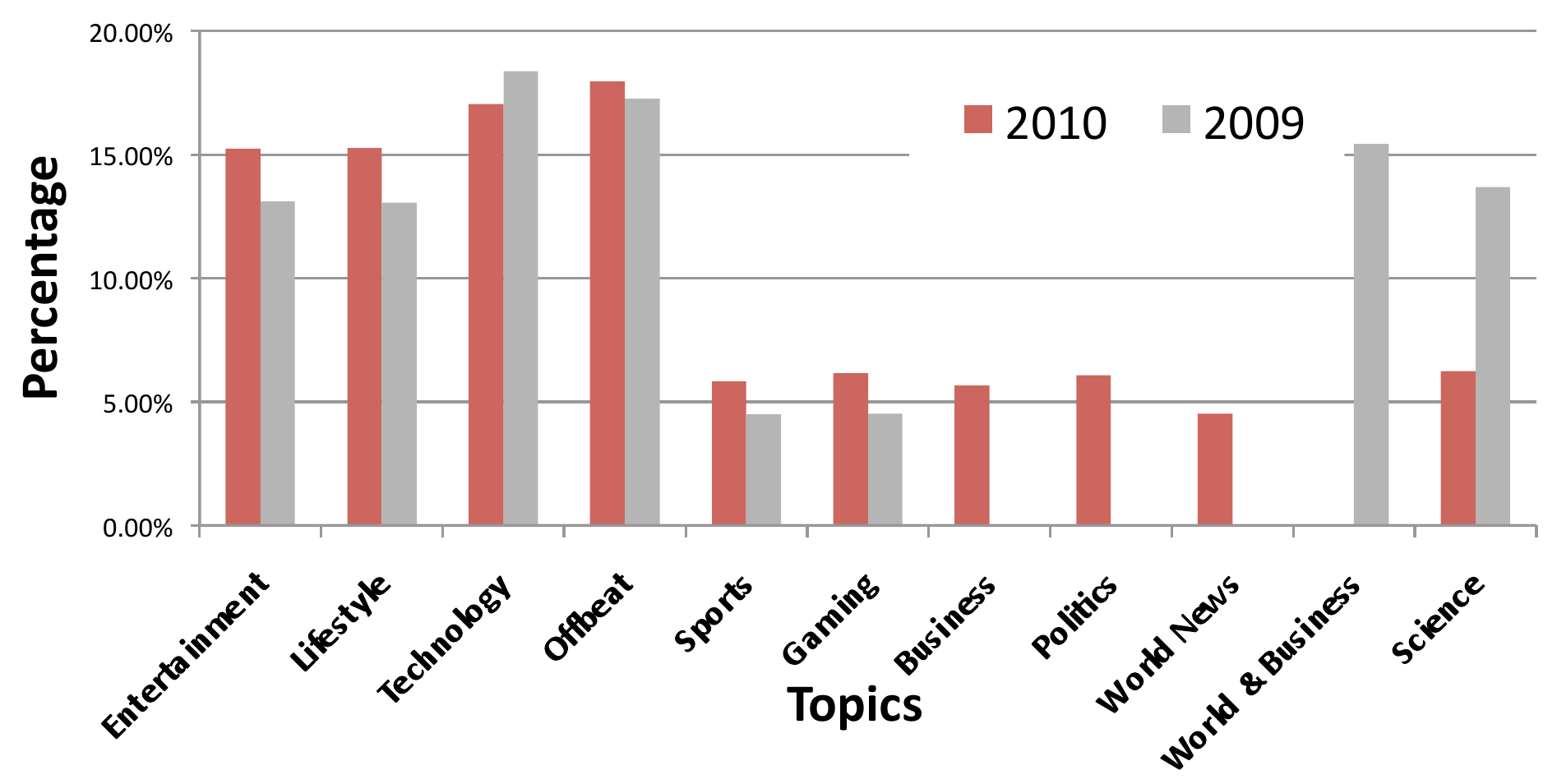}
\end{tabular}
\end{center}
\caption{Topic distribution in 2009 and 2010 dataset.}
\label{fig:topicPercentage}
\end{figure}

We use the topics Digg assigns to each story to represent its content. Figure~\ref{fig:topicPercentage} shows the distribution of Digg-assigned topics in our data sets, that is the percentage of stories assigned to each topic. In both data sets, ``Offbeat,'' ``Entertainment,'' ``Lifestyle'' and  ``Technology'' were the most popular topics, while ``Sports'' and ``Gaming'' were the least popular topics. Overall, there is no dominant topic in either dataset and the popularity ranking of different topics are almost identical. The topics assigned to stories by Digg provide useful evidence for identifying user's topic preferences. We represent user $u_i$'s topic interest vector $\theta_{i}$ by computing the fraction of votes he made on each  topic.

\paragraph{Topic Diversity}
The variance of topics to which a user is exposed to by his friends has important implication for modeling information in a social network. Topic diversity of a user's network neighborhood measures the variance of friends' topic interests: when most of friends have orthogonal interests, topic diversity will be high, whereas when most of friends have similar topic interests it will be low.

Diversity can be captured in several different ways. Aral \& Van Alstyne~\cite{Aral11,aral2012anatomy} defined topic diversity as the average cosine distance of friends' topic vectors and their mean topic vector aggregated over all friends. Based on our experiments, Aral \& Van Alstyne's measurement is not able to capture topic diversity correctly for users with the same mean (based on friend topic vectors) but different number of friends.
Instead, we define a user $u_i$'s topic interest vector $\theta_i$ in terms of the Digg-defined categories. Each  component of $\theta_i$ represents the fraction of all votes made by $u_i$ on stories belonging to that category. Then, we define topic diversity of a user's network by averaging pair-wise cosine distances of friends' topic interest vectors.
\begin{equation}
\begin{aligned}
TD_{i}= \frac{ \sum_{j=1}^{N^{frd}_i} \sum_{k=1}^{N^{frd}_i} (1-Cos(\theta_{j},\theta_{k}))}{S^{2}_{i}}
 \end{aligned}
\end{equation}

\paragraph{Novel Information}
Total amount of novel information is another important measure of information content of networks. In many social media services, the same message or a piece of information can be recommended multiple times by multiple friends. Since most social media services provide a unique identifier for each message  (e.g., original tweet id on Twitter or story id on Digg), we can measure the amount of novel information that a user is exposed to during time period $\Delta T$ by counting the number of distinct messages, or stories on Digg, to which user's friends expose her. Following Aral \& Van Alstyne, we refer to this quantity as $NRI_i$, or \emph{non-redundant information}, although in Aral \& Van Alstyne's studies, this quantity was not measured directly but derived from topic diversity and friend activity. In addition to the amount of novel information, we can also measure the novel information rate in a user's social network as $R_{i} = NRI_{i}/I_{i}$.
Of the total volume of novel information ($NRI_i$)  $u_i$  is exposed to through friends' recommendation activities, $u_i$ adopts a subset $O_{i}^a$ based on the topics or popularity of information. We measure the novel information adoption rate by $NAR_{i} = O_{i}^a/NRI_{i}$.

\paragraph{Novel Information Potential}
In social media, user's access to novel information is mainly determined by the activities of their friends. We introduce a new variable to define the volume of novel information that a user could potentially be exposed to if his friends adopted all the information they themselves were exposed to. We measure $NRI^{frds}_i$, the potential amount of  novel information $u_i$ could access, by counting the number of distinct stories that all friends of $u_i$ are exposed to.
While the friends of $u_i$ have access to $NRI^{frds}_i$ novel information, they adopt a subset of this information based on their interests, exposing $u_i$ to $NRI_{i}$  novel information. We measure friend novel information adoption rate by $FNAR_{i} = NRI_{i}/NRI^{frds}_i$.

\section{How Structure and Activity\\ Shape Information Access}
\label{sec:networkstructure}

Sociologists have long noted a relationship between the structure of a social network and the frequency and intensity of interactions between two people, what sociologists call the strength of a tie. Granovetter~\cite{granovetter1973} argued that social tie strength can be estimated from local network structure of the two people, specifically, the number of common neighbors they have. Subsequently, a study of a massive mobile phone network established a correlation between frequency and duration of phone calls (one measure of a strength of a tie) and the fractions of common neighbors callers have~\cite{onnela2007structure}. This makes sense; since close friends not only interact with each other frequently and intensely, but also are likely to revolve in the same social circles, therefore, share common friends. On the other hand, weak ties, or acquaintances, are pairs people who interact infrequently. They are likely to come from different social circles or communities, and therefore, not share common friends.

The relationship between tie strength and access to novel information is more subtle. Though weak ties deliver novel information~\cite{granovetter1973}, for example, new job prospects, since the volume of communication along these ties is low, so is their potential to deliver novel information. This was confirmed by Aral \& Van Alstyne's analysis~\cite{Aral11,aral2012anatomy} of email communication within a corporate recruiting firm. They showed that structurally diverse networks provide access to diverse and novel information, though the positive effects of structural diversity are offset by lower volumes of communication (bandwidth),  what they call ``diversity--bandwidth trade-off.''

To date, little is known about how these factors operate in online social networks and how they compare to real-world and email networks. Ties in online social networks, including Digg, are often non-reciprocal, with users sharing messages with both friends they know in real life and strangers. We explore the questions about how users can broaden their access to information by controlling their position within the network and their activity level.

\subsection{Access to Information}

In the study of email communication within a corporate recruiting firm, Aral \& Van Alstyne observed that both the  total volume of novel information ($NRI_i$) flowing to recruiters and its diversity  ($TD_i$) increased with their network size, network diversity and channel bandwidth (the number of emails they received along each tie).
We tested whether the same conclusions hold for the online social network of Digg. Specifically, whether larger ($S_i$), structurally diverse networks ($ND_i$) or high friend activity ($B_i$) are likely to deliver more novel information ($NRI_i$) and topically diverse information ($TD_i$).

\subsubsection{Effect of Network Size}
\begin{figure}[tbh]
\begin{center}
\begin{tabular}{c}
\\
\includegraphics[width=0.9\linewidth]{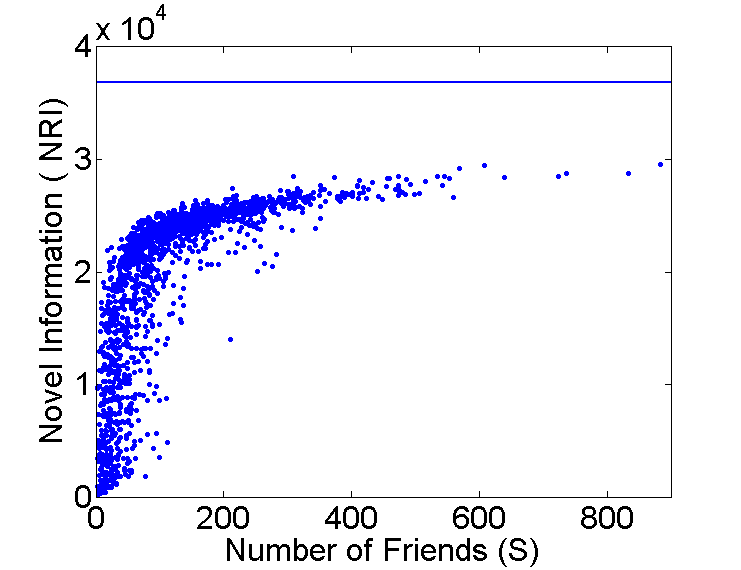} \\
\end{tabular}
\end{center}
\caption{Amount of novel information ($NRI_i$) a user is exposed to as a function of the number of active friends ($S_i$) in the 2010 Digg data set. The line represents the total number of distinct stories in the data set. }
\label{fig:nobodyaccess}
\end{figure}

One of simplest ways users can control their position within a network is by adding friends. But, does having more friends improve access to information in online social networks? We study how the volume of novel information a user can access, which we measure by the number of distinct stories to which the user is exposed by friends on Digg, varies with the number of friends.  Figure~\ref{fig:nobodyaccess} shows the volume of novel information ($NRI$) users can access as a function of the number of friends ($S$). The amount of novel information increases as users add more friends, but saturates quickly. 
Surprisingly, no single user had access to all the information available in the network (shown as a line in Figure~\ref{fig:nobodyaccess}). The highest number of distinct stories any user was exposed to in the 2010 Digg dataset was 29,558, or 80\% of the total (36,883 distinct stories). It appears that adding more friends in an online social network improves access to novel information, but very quickly, after about 100 friends, it becomes counterproductive, since doubling the number of friends raises the volume of novel information only a few percentage points.

\subsubsection{Effect of User Activity}

\begin{figure}[tbh]
\begin{center}
\begin{tabular}{c}
\\
\includegraphics[width=0.9\linewidth]{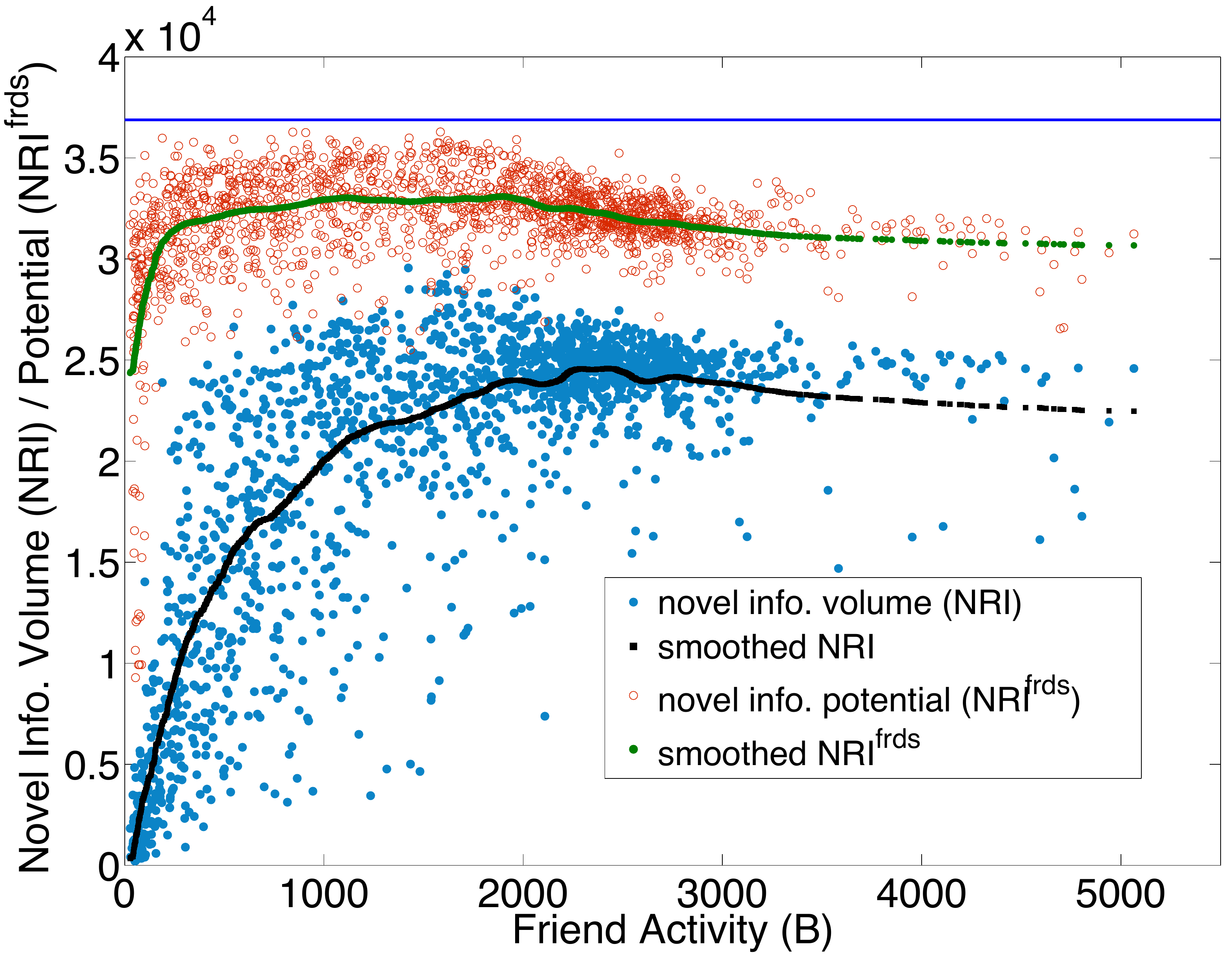} \\
(a) \\
\includegraphics[width=0.9\linewidth]{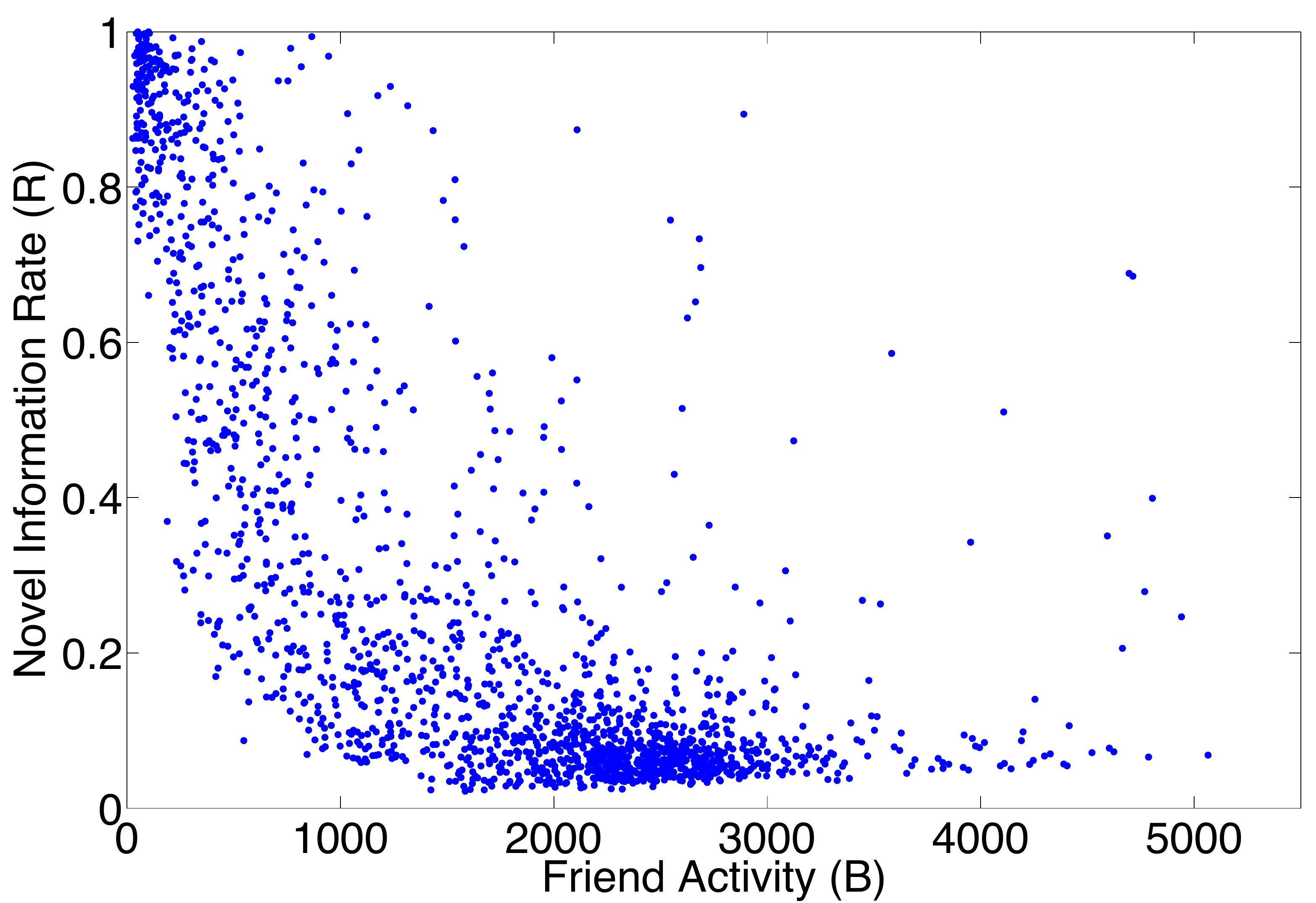}\\
 (b) \\
\end{tabular}
\end{center}
\caption{Novel information in a user's network in the 2010 Digg data set. (a) The total amount of novel information that user's friends ($NRI^{frds}$) and the user ($NRI$) are exposed to as a function of average friend activity (or channel bandwidth $B$). Solid symbols show smoothed data, and the line represents the total amount of information in the network (number of distinct stories in the data set). (b) Novel information rate as a function of  friend activity.}
\label{fig:activity}
\end{figure}

In addition to creating new social links, a user can choose to link to more active users in order to improve his access to information in a social network. Does having active friends, i.e., friends who recommend many stories, lead to greater access to novel information? \figref{fig:activity} shows the relationship between the volume of novel information in a user's network as a function of  average friend activity (referred to as channel bandwidth by Aral \& Van Alstyne).   \figref{fig:activity} (a) shows the amount of novel information that user's friends are exposed to ($NRI^{frds}_i$). The solid line represents the total amount of information in the network, i.e., distinct stories in the data set. Potential amount of novel information rises quickly as a function of friend activity, approaching near-maximum. However, the amount of novel information to which the user is exposed is just a fraction of this maximum, as shown in \figref{fig:activity} (a).
Interestingly, the amount of potential  novel information and novel information available to the user both decrease as friend activity grows past 2000.
Our results indicate that while linking to more active users does initially improve access novel information in a social network, after a certain point, higher friend activity no longer increases the amount of novel information available to the user, but may even slightly suppress it.

\figref{fig:activity} (b) shows the rate at which users receive novel information, i.e., the fraction of novel information in their information stream, as a function of the average activity of their friends (channel bandwidth $B_i$). The figure clearly shows that as friends become more active, by voting for more stories, the fraction of novel information in the user's social stream drops precipitously. As we show later, this is due to higher redundancy of incoming information.
In online social networks, friends activity is an important factor in deciding the extent and the amount of novel information available to the user.


\subsubsection{Effect of Network Structure: the ``Diversity--Bandwidth Trade-off''}

\begin{figure}[]
\begin{center}
\begin{tabular}{c}
\includegraphics[width=1.0\linewidth]{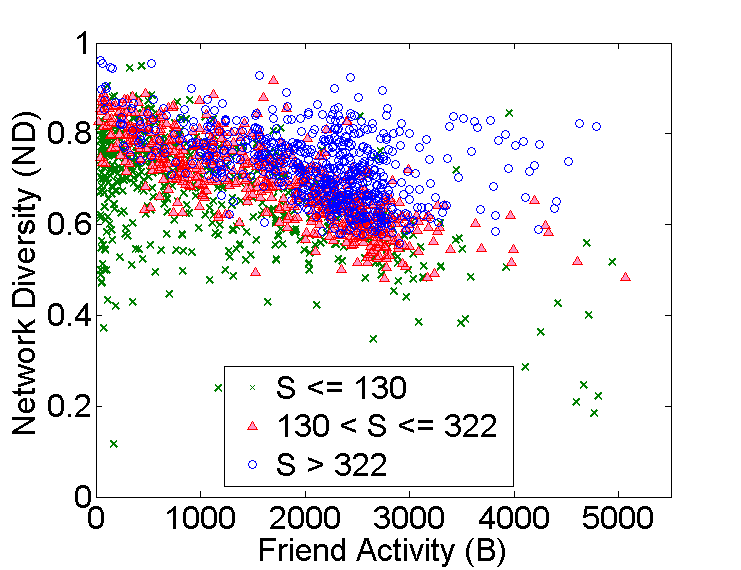}
\end{tabular}
\end{center}
\caption{Scatterplot showing network diversity vs average friend activity (channel bandwidth) for Digg users who are divided into three populations based on the number of friends in the 2010 Digg data set. The plot demonstrates the diversity-bandwidth trade-off.}
\label{fig:dbtradeoffs}
\end{figure}

Next, we study the interplay between network structure and user activity and their impact on access to information. Aral \& Van Alstyne demonstrated that while structurally diverse networks provide greater access to information, their benefits are offset by lower rate of communication along structurally diverse ties. Due to such ``diversity--bandwidth trade-off,'' people can increase their exposure to topically diverse and novel information either by placing themselves in structurally diverse network positions or by linking to people with higher bandwidth who will communicate with them more frequently.

We examine whether ``diversity--bandwidth trade-off'' exists on Digg. Digg users can increase their ``bandwidth'' by linking to friends who vote for more stories.
However, as friends' activity increases, it worsens the user's cognitive load, or the volume of incoming information the user has to process.
We divide users into different populations based on the total volume of incoming information,  which is,  on average, proportional to the number of active friends $S_i$ they have.  Figure~\ref{fig:dbtradeoffs} shows the relationship between network diversity $ND$ and average friend activity (or channel bandwidth) $B$ for each user, where users are broken into three populations: those with more than 322 friends, between 131 and 322 active friends, and 130 or fewer active friends. The thresholds were chosen to produce equal size populations. The correlation between network diversity and average friend activity for the three populations are  -0.54 (p<.01),  -0.58 (p<.01) and -0.50 (p<.01) respectively. Overall (over all populations of users), there is still a strong negative relationship (-0.47, p< .01) between network diversity $ND_i$ and bandwidth $B_i$,  confirming the ``diversity--bandwidth trade-off''~\cite{Aral11}: users who place themselves into positions of greater network diversity within the Digg follower graph on average receive fewer story recommendations from friends than users who place themselves into positions of smaller network diversity. For 2009 data set, we also divided users into three populations: those with more than 87 friends, between 26 and 87 active friends, and 25 or fewer active friends. The correlation between network diversity and average friend activity are  -0.54 (p<.01),  -0.59 (p<.01) and -0.03 (p<.01) respectively and over all populations of users in 2009 data set, the correlation is -0.13 (p<.01). The differences mainly coming from incomplete history about users' activities in 2009 data set, since it only contains subset of users' behaviors on front page stories, while we have the complete users' voting history in the 2010 data set.

In both 2009 and 2010 Digg data set, users in the greater network diversity within the follower graph on average receive fewer story recommendations from friends than users who place themselves into positions of smaller network diversity. We observed that users connected by strong ties are more active, recommending more stories than those users who are connected by weak ties. Similarly, as users' networks become more diverse, friends' activities contract.

\begin{figure}[tbh]
\begin{center}
\begin{tabular}{c}
\includegraphics[width=0.9\linewidth]{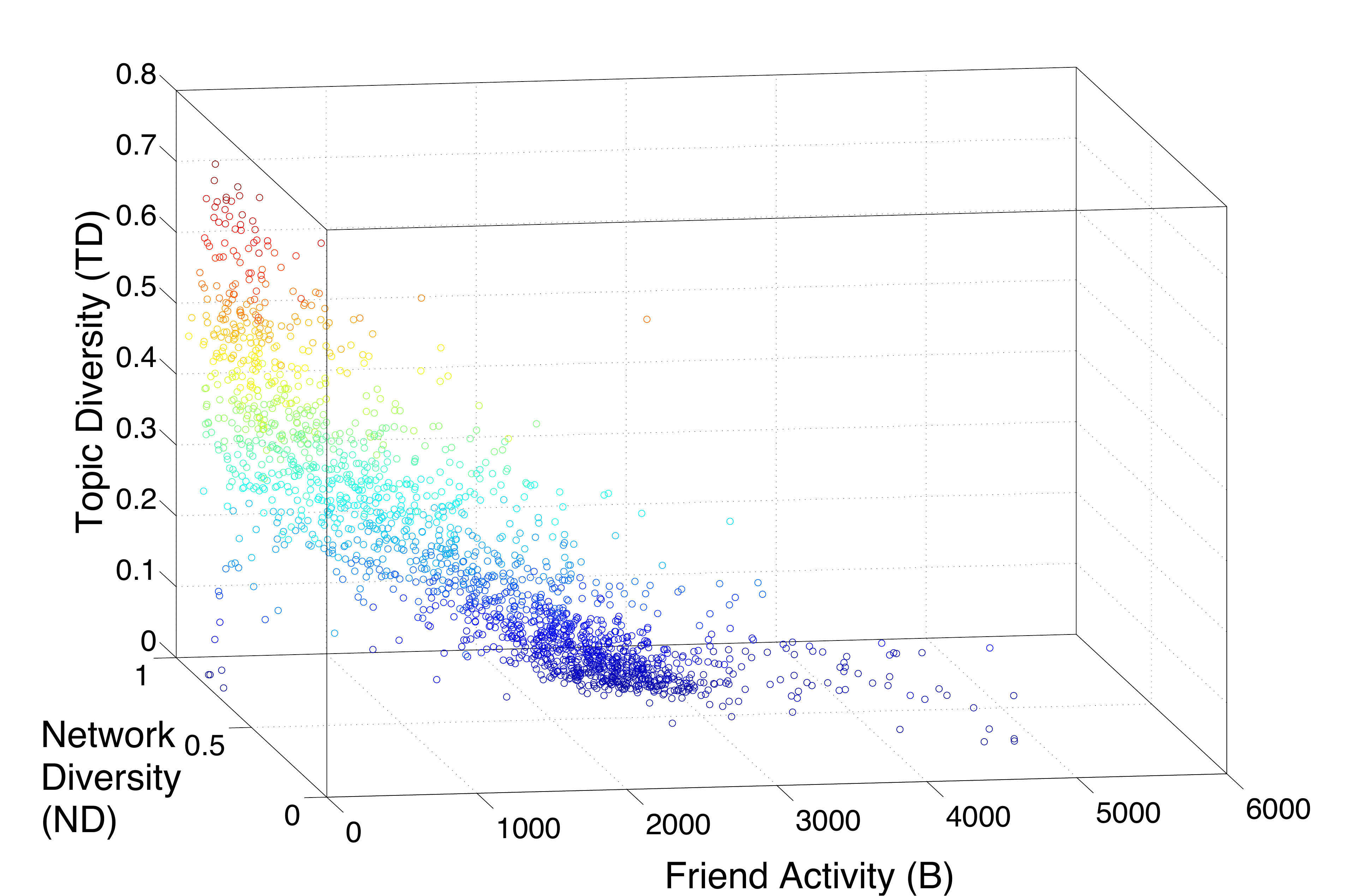}\\
(a)\\
\includegraphics[width=0.9\linewidth]{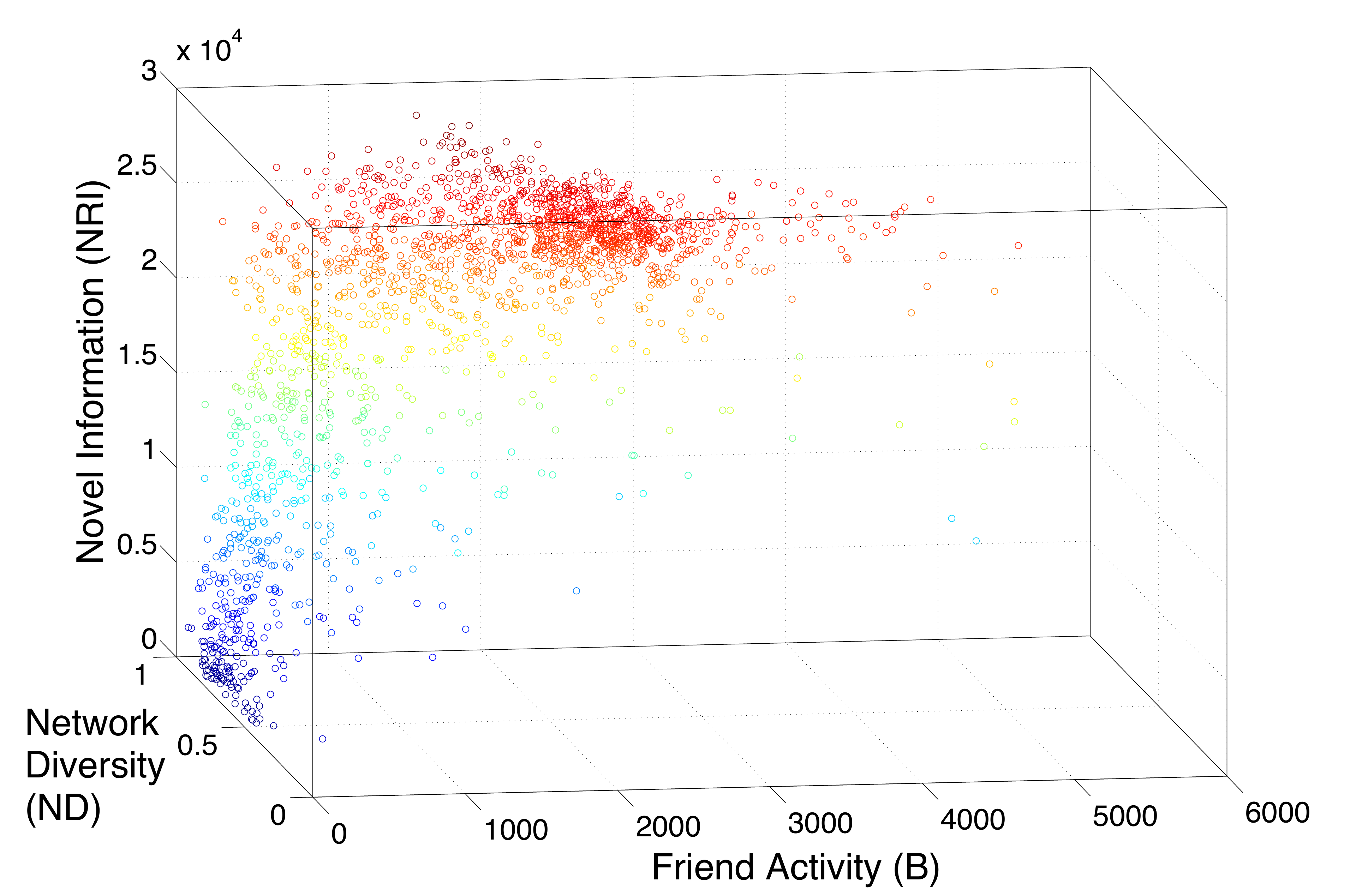}\\
(b)
\end{tabular}
\end{center}
\caption{(a) Topical diversity ($TD$) and (b) novelty ($NRI$) of information to which Digg users are exposed as a function of their network diversity ($ND$) and average friend activity (or channel bandwidth $B$) in the 2010 Digg data set.}
\label{fig:DB_TD_NRI}
\end{figure}

\begin{table}
\center
\begin{tabular}{|c|c|c|c|c|c|c|}
\cline{1-3} \cline{5-7}
\textbf{2009} &\textbf{   NRI }  & \textbf{TD} &&\textbf{2010} &\textbf{   NRI }  & \textbf{TD}\\
\cline{1-3} \cline{5-7}
\textbf{B} &  0.04**& -0.11** & &\textbf{B}&0.69** & -0.83**\\
\textbf{ND} & -0.09** & 0.48** & &\textbf{ND}& -0.15**& 0.41**\\
\cline{1-3} \cline{5-7}
\end{tabular}
\center
\caption{Pairwise correlations between variables in the 2009 and 2010 Digg data sets. Note that asterisk (**) shows statistically significant correlation with p<.01.}
\label{tbl:pairwisecorr}
\end{table}

Figure~\ref{fig:DB_TD_NRI} shows how much (a) topically diverse information ($TD$) and (b) novel information ($NRI$) Digg users are exposed to as a function of  their position in the network (network diversity $ND$) and friend activity ($B$). Users whose friends are more active can access more novel information ($NRI$), whereas users in positions of higher network diversity can access more topically diverse information ($TD$). This is in contrast to the findings of Aral \& Van Alstyne, which demonstrated that users could increase both the topic diversity and amount of non-redundant (novel) information they are exposed to by increasing either their network diversity or channel bandwidth. On Digg, on the other hand, users who place themselves in position of high structural diversity can access more topically diverse information (correlation between $ND_i$ and $TD_i$ was 0.41 (p<.01)), rather than more novel information. There was a strong negative relationship (in Table~\ref{tbl:pairwisecorr}) between $B_i$ and $TD_i$ that shows users in a strongly tied network have similar topical interests. Based on detailed investigation, the strong negative relationship is not because of friends' uniform preferences over a variety of topics but because of high similarities between friends' topic preferences in highly clustered network. Intensifying friends' activity, on the other hand, led to more novel information (correlation between  $B_i$ and $NRI_i$ was 0.69 (p<.01)), but less topically diverse information.
These results demonstrate that activity is an important feature for accessing novel information in online social networks, while structural diversity can be used to get access to more topically diverse information.

\begin{figure}[tbh]
\begin{center}
\begin{tabular}{c}
\includegraphics[width=1.0\linewidth]{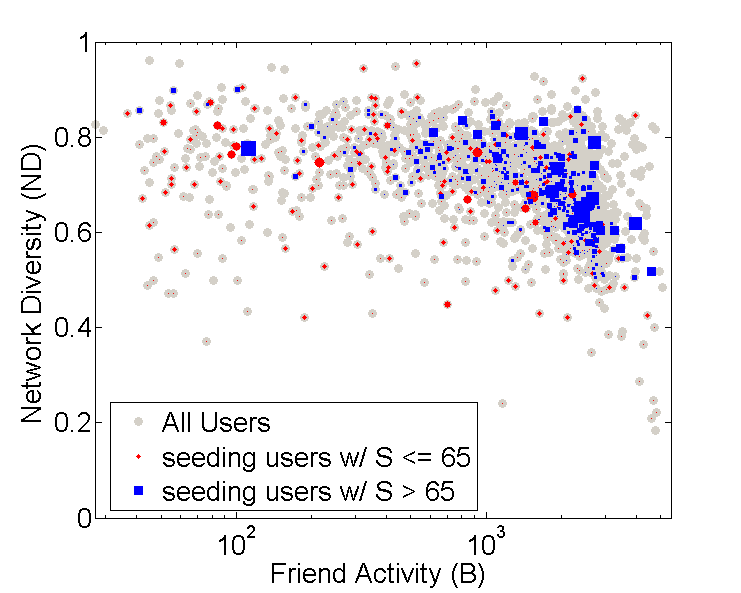}
\end{tabular}
\end{center}
\caption{Amount of new information injected into the network by users (colored symbols) in different positions of network diversity (ND) and friend activity (B). Symbol size represents the relative number of seeded stories. Seeding users are divided into classes based on the number of friends.}
\label{fig:seeding}
\end{figure}

Last, we examine the characteristics of users who inject new information into their networks by voting for stories they found outside of their friends' recommendations, e.g., on the Web or on other sections of Digg.
Figure~\ref{fig:seeding} shows the network diversity vs friend activity plot with colored dots representing users who introduce, or seed, new stories in their network. We divide these users into two classes based on the number of friends they have. The size of the symbol represents the relative number of seeded stories (difference between the total votes made by $u_i$ and those adopted through friends' recommendations). The x-axis is shown in log-scale to highlight the differences between classes of users.
Users with many friends (blue symbols) who are very active (high $B$) inject relatively more new stories into their network than users with many, but less active friends. These users are also in less structurally diverse positions, i.e., they are more strongly tied to their friends. At first, it seems counterintuitive that these users, who already receive many recommendations, would take the time to look for new information. These could be the dedicated top users, who consider it their responsibility to look for new stories to post on Digg, or users who are so overwhelmed with the quantity and redundancy of their friends' recommendations, that they choose to find new information on their own.
Users with few friends (red symbols) also tend to have less active friends (lower $B$). These users inject more information into their network when their network diversity is high, or as shown earlier, their friends have diverse interests. Such users cannot rely on their network to expose them to interesting information; instead, they seek it out themselves by seeding new stories.


\subsection{Bottlenecks to Information Access}



\begin{figure}[tbh]
\begin{center}
\begin{tabular}{c}
\includegraphics[width=0.95\linewidth]{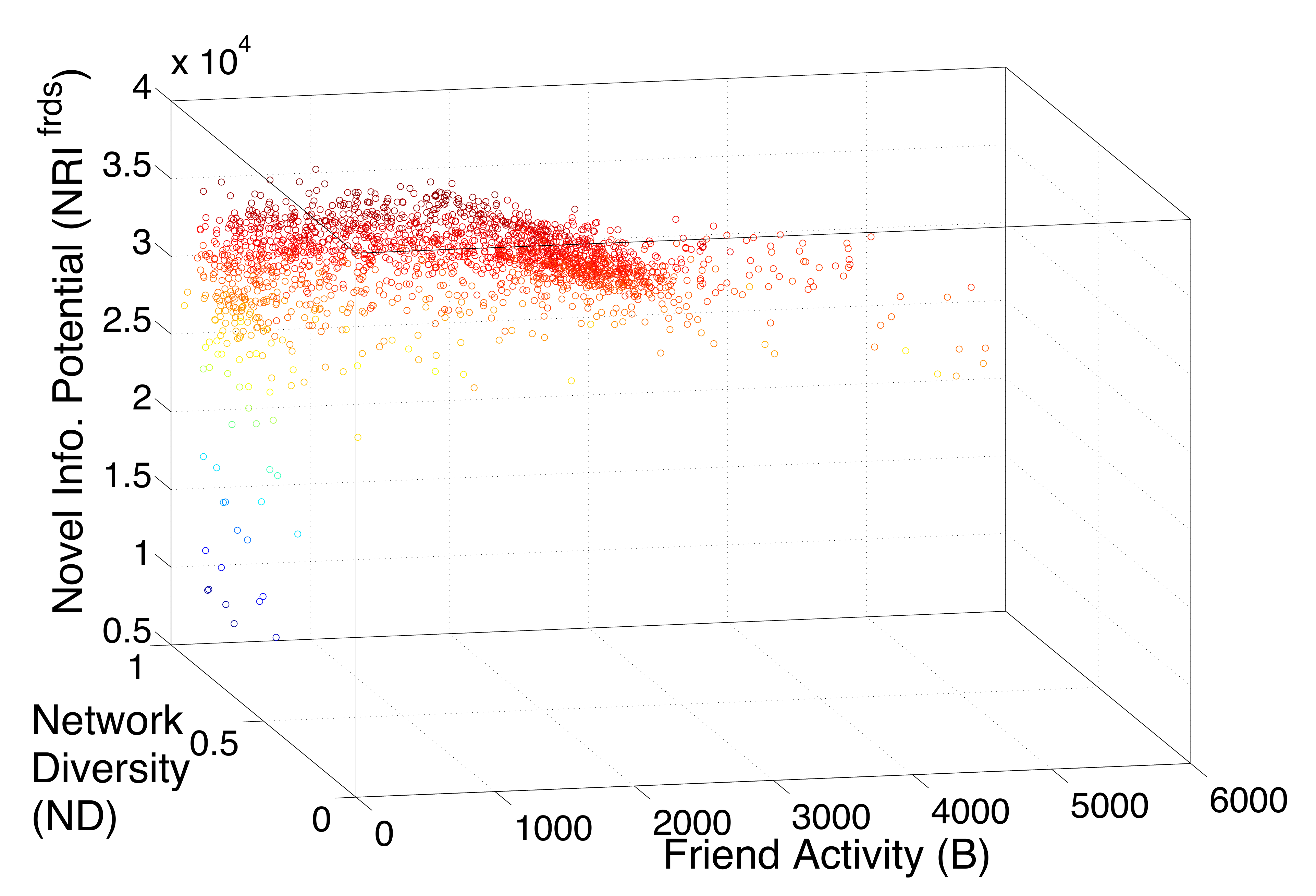}\\
(a)\\
\includegraphics[width=0.95\linewidth]{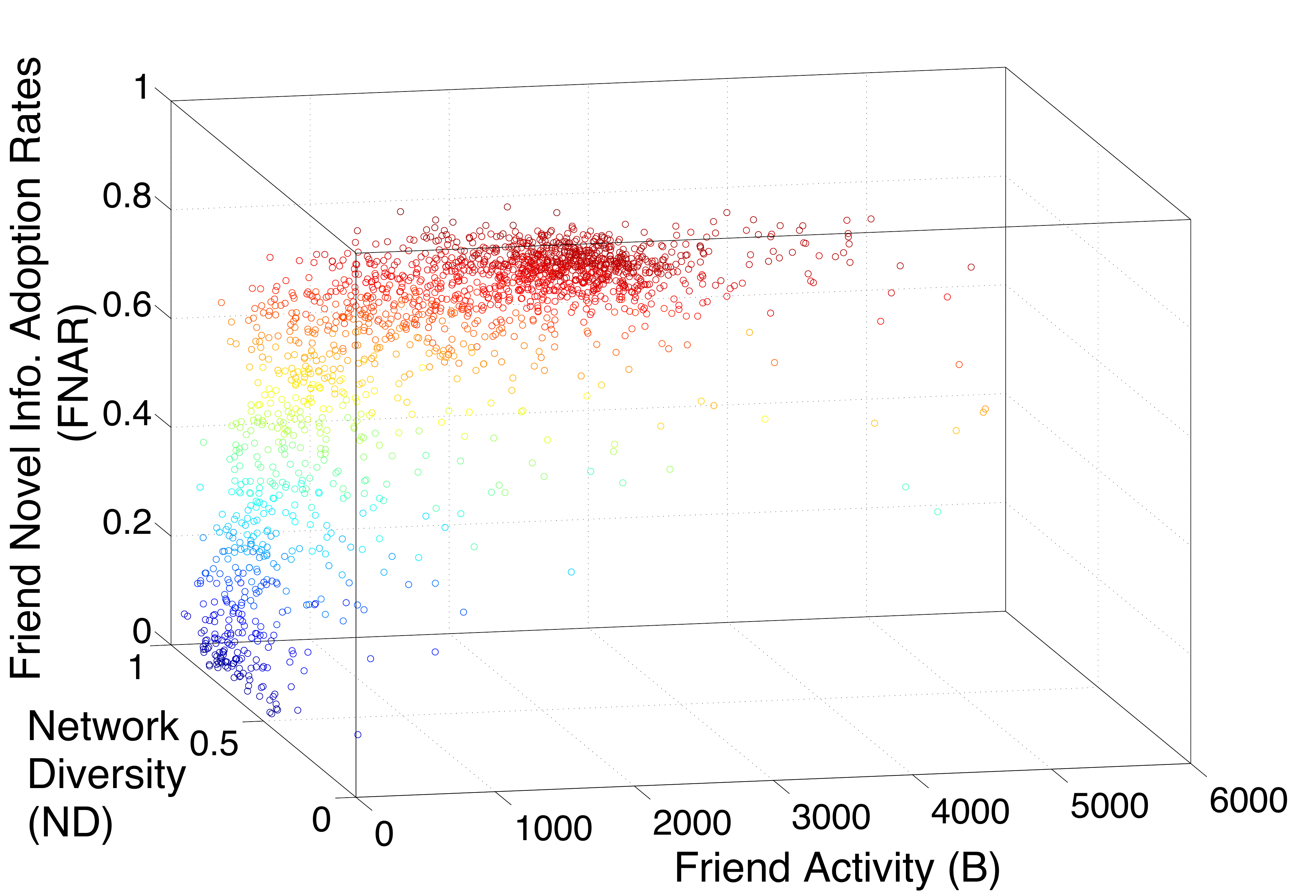}\\
(b)
\end{tabular}
\end{center}
\caption{(a) Total amount of novel information that user's friends are exposed to ($NRI^{frds}_i$) as a function of user's network diversity ($ND_i$) and friend activity ($B_i$)  in the 2010 Digg data set.  (b) Fraction of novel information adopted ($FNAR_i$) by friends ($NRI_i/NRI^{frds}_i$) as a function of network diversity and friend activity in the 2010 Digg data set.}
\label{fig:NRI_frd}
\end{figure}

Why do users in positions of high network diversity receive less novel information even though they are connected to more topically diverse friends? To answer this question, we measured $NRI^{frds}_i$, the total amount of novel information that friends of user $u_i$ are exposed to. \figref{fig:NRI_frd} (a) shows that this quantity depends both on network diversity ($ND$) and friend activity ($B$). In most cases, friends are collectively exposed to a large quantity of novel information (also demonstrated by \figref{fig:activity} (a)), almost all of the 36,883 distinct stories in the 2010 Digg data set. Although most of the users could potentially be exposed to nearly all of the information in the network, in fact, as shown in ~\figref{fig:activity}, they receive far less novel information. We get some insight into this puzzle from \figref{fig:NRI_frd} (b), which shows the friends' novel information adoption rate, i.e., the fraction of stories in their stream friends voted for, as a function of the user's network position and friend activity rates. Friends of users in positions of high network diversity fail to adopt most of the novel information they are exposed to. However, users with highly active friends (high channel bandwidth $B$ region) are exposed to more novel information because their friends adopt it at a higher rate.

This could explain the difference between our study and the findings of Aral \& Van Alstyne. In their study, users could increase their access to diverse and non-redundant (novel)  information by increasing their network diversity or channel bandwidth. In our study, however, users in positions of high friend activity (high channel bandwidth) increase their access to novel information since their friends adopt a large portion of the novel information that they themselves are exposed to. Users in  positions of high network diversity are exposed to more diverse information, but since their friends have interests that are different from their own, they do not adopt much of the information they are exposed to. In addition, in Aral \& Van Alstyne's study, novelty and diversity were not independent variables: non-redundant information (novelty) was the product of topic diversity and channel bandwidth. Hence, it may not be surprising that both were correlated highly with network diversity and channel bandwidth.  We treat these variables as independent variables: while topic diversity of a network neighborhood is measured based on the Digg topic assignments of the stories friends adopted, non-redundant, or novel, information is measured simply by the number of distinct stories friends adopted.

Our study demonstrates that friends in structurally diverse positions and positions of high activities constrain user's information exposure in different ways. Increasing friend activity affects novel information access, while increasing network diversity provides access to more topically diverse friends, but not the other way around.


\begin{figure}[tbh]
\begin{center}
\begin{tabular}{c}
\includegraphics[width=0.9\linewidth]{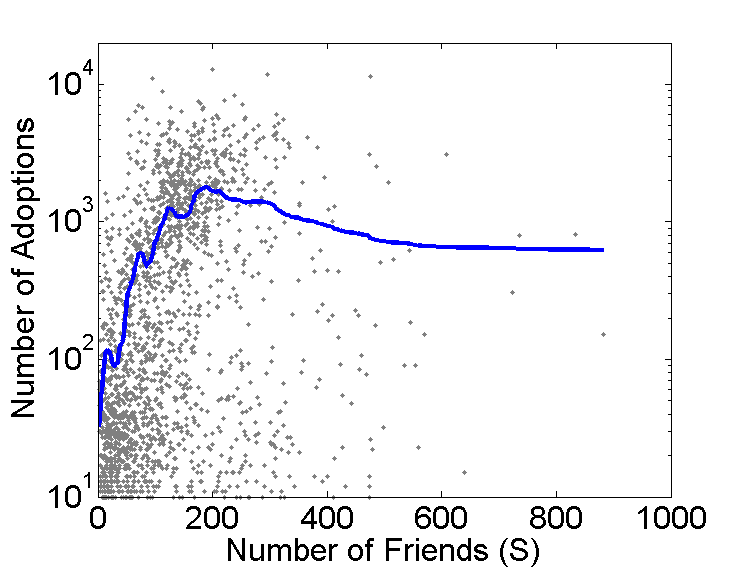}
\end{tabular}
\end{center}
\caption{Number of stories adopted by the user as a function of the number of active friends  ($S$) in the 2010 Digg data set.}
\label{fig:uB_S}
\end{figure}

Other factors that affect access to information are the cognitive constraints that limit user's ability to process incoming information. These cognitive constraints are similar to those that limit the number of stable social relationships a human can maintain~\cite{Dunbar}. In the social media domain, they have been shown to limit the number of conversations users engage in~\cite{Goncalves11} and limit the spread of information~\cite{Hodas12socialcom}. We believe that cognitive constraints also play a role in access to information on networks. As the volume of incoming information increases, users compensate by increasing their activity to cope with the greater amount of new information. This works, but up to a point. Unfortunately, we cannot directly measure how much information users process, e.g., how many stories they read, but on average this quantity should be proportional to the number of recommended stories they vote for. We call this quantity user activity $uB_i$. Figure~\ref{fig:uB_S} shows user activity  as a function of the number of active friends the user has. While user activity initially increases with the number of friends, it reaches a maximum around 200 friends and then decreases somewhat. The cognitive constraints prevent users from keeping up with the rising volume of new information friends expose him to. This, in turn, limits the amount of new information to which they expose their followers. This effect is more dramatic for users whose friends have topically diverse interests. These friends tend to vote on fewer recommended stories, either because they lack interest in processing recommended information, or because they are less willing to devote the greater cognitive resources required to process this diverse information.
Understanding the complex interplay between cognitive constraints and network structure is the subject of our ongoing research.


\section{Related Work}

Sociologists have long argued that network structure determines the intensity of interactions between  people and affects information diffusion in the network. The theoretical arguments known as ``the strength of weak ties'' were proposed by Mark Granovetter~\cite{granovetter1973}, who showed that social tie strength can be estimated from local network structure. In the same paper, he linked tie strength to information access in a network. Specifically, he argued that weak ties provide users with access to novel information. 
The theory of weak ties has been verified by many empirical studies~\cite{uzzi1997social,allen2003managing,reagans2001networks,reagans2003network,onnela2007structure}, including studies of job search~\cite{granovetter1973}, business relations~\cite{coleman1988social,Aral11}, inter-business~\cite{uzzi1996sources}, social capital~\cite{coleman1988social}.
Burt~\cite{Burt95,burt2005brokerage} argued that weak ties act as bridges between different communities and enable ``brokers'' to leverage diverse sources to access novel information. Empirical studies of mobile phone~\cite{onnela2007structure} and email communication~\cite{iribarren2011affinity,Aral11} has offered support for brokerage theory.

Several studies have confirmed the role of weak ties as bridges between different communities~\cite{centola2010spread,journals11074009} though their impact on information diffusion in online networks is still debated~\cite{centola2007complex,journals10013181}. To the best of our knowledge, out study is the first to look at the relationship between network structure and access to novel information in online social networks.

Aral \& Van Alstyne~\cite{Aral11} examined the relationship between weak ties, structural diversity and access to diverse and novel information. Their study of email communication within an organization demonstrate a trade-off between network diversity (brokerage positions that link different communities) and channel bandwidth (tie strength) in access to both diverse and non-redundant (novel) information. In a follow-up study they demonstrated importance of network position in maximizing information diversity and novelty~\cite{aral2012anatomy}. 
In contrast, our study demonstrates that users with very active friends (high channel bandwidth) increase their access to novel information since their friends adopt a large portion of the novel information that they themselves are exposed to. Users in  positions of high network diversity are exposed to more diverse information, but since their friends have interests that are different from their own, they do not adopt much of the information they are exposed to. Moreover, with more freedom of users' activities in online social network, we studied the contribution of users activities to access to diverse and novel information as well as ``diversity--bandwidth trade-off.''

Homophily, which refers to the tendency of similar individuals to link to each other, is a strong organizing principle of social networks. Numerous studies found that people tend to be friends with other who belong to the same socio-economic class~\cite{Feld1981,mcpherson2001birds}, and they tend to follow others in social media who have similar interests~\cite{Kang12aaai}. In the context of information exchange on networks, this means that content users are exposed to in social media depends on their position in the network. Users embedded within a community of strongly tied individuals are likely to share information on topics that other community members are interested in, while users in brokerage  positions that bridge different communities receive information on more diverse topics. In this paper, we show that the variance of topics to which a user is exposed to by his friends is highly related to the structure diversity of social network.

Recently, researchers recognized that cognitive constraints are important to defining social interactions online and in the real world. The number of social relationships that people can maintain is limited to about 150~\cite{Dunbar03}. This is similar to the limit of the number of conversation partners that Twitter users have~\cite{Goncalves11}. Cognitive constraints, specifically, divided attention, was also shown to limit the spread of information on Twitter~\cite{Hodas12socialcom}. We find a dependence of user activity on network structure that mirrors those imposed by cognitive constraints. We observe that user's activity rate initially increases with the number of friends, until reaching a maximum around 200 friends and then decreases somewhat. The cognitive limits constraints the potential exposures to new information as users have too much information recommended by many friends. Further we argued that the limited activities in high diverse network is either because their lack of interest in processing recommended information, or because they are less willing to devote the greater cognitive resources required to process this diverse information.


\section{Conclusion}

We used data from the social news aggregator Digg to investigate the relationship between the structure of the follower graph, user activity, and access to information in social media.  We showed the amount of novel information a user is exposed to increases as she adds more friends, but saturates quickly. Similarly, linking to friend who are more active improves access to novel information, but as the redundancy increases, higher friend activity can no longer increase the amount of novel information accessible to the user.
In addition, we validated the ``diversity--bandwidth trade-off'' in online social media. In two different data sets, users in positions of greater network diversity in the follower graph on average receive fewer story recommendations from friends than users who place themselves into positions of high friend activity (high bandwidth). Users in positions of higher network diversity can access more topically diverse information while users whose friends are more active can access more novel information.  Increasing friend activity affects novel information access, while increasing network diversity provides access to more topically diverse friends, but not the other way around. So to access more novel information the volume of communication along the tie is more important than the network structure. This is in contrast to the findings of Aral \& Van Alstyne, who demonstrated that users could increase both the topic diversity and amount of non-redundant (novel) information they are exposed to by increasing either their network diversity or channel bandwidth in the email communication.

Cognitive constraints, e.g., limited attention, is an important psychological factor that limits human activity. Our analysis of Digg suggests that user's activity creates an ``information bottleneck,'' blocking potential access to novel information to their followers. Since user's network diversity is highly related to topic diversity, the mechanisms that adopted novel information become even more important in diverse network. 
Understanding the complex interplay between cognitive constraints and network structure is the subject of our ongoing research. 


\section*{Acknowledgment}
This material is based upon work supported by  the Air Force Office of Scientific Research under Contract Nos. FA9550-10-1-0102 and FA9550-10-1-0569, and by DARPA under Contract No. W911NF-12-1-0034.

\bibliographystyle{abbrv}
\bibliography{reference}
\end{document}